\documentclass[runningheads]{llncs}
\usepackage[T1]{fontenc}
\usepackage{graphicx}
\usepackage{amsmath}
\usepackage{multirow}
\begin{document}
\title{Benchmarking Self-Supervised Models for Cardiac Ultrasound View Classification}
\author{Youssef Megahed \inst{1,2}\orcidID{0009--0004--2595--5468} \and
Salma I. Megahed \inst{11}\orcidID{0009--0008--7763--1288} \and
Robin Ducharme \inst{3}\orcidID{0000--0002--5665--7429} \and
Inok Lee \inst{3} \and
Adrian D. C. Chan \inst{1}\orcidID{0000--0002--2111--247X} \and
Mark C. Walker \inst{3,4,5,6,7}\orcidID{0000--0001--8974--4548} \and
Steven Hawken \inst{1,2,5}\orcidID{0000--0002--3341--9022}}
\authorrunning{Y. Megahed et al.}
\institute{Department of Systems and Computer Engineering, Carleton University, Ottawa, Ontario, Canada \and
Department of Methodological and Implementation Research, Ottawa Hospital Research Institute, Ottawa, Ontario, Canada \and
Department of Acute Care Research, Ottawa Hospital Research Institute, Ottawa, Ontario, Canada \and
Department of Obstetrics and Gynecology, University of Ottawa, Ottawa, Ontario, Canada \and
School of Epidemiology and Public Health, University of Ottawa, Ottawa, Ontario, Canada \and
Department of Obstetrics, Gynecology \& Newborn Care, The Ottawa Hospital, Ottawa, Ontario, Canada \and
International and Global Health Office, University of Ottawa, Ottawa, Ontario, Canada \and
College of Medicine, Alfaisal University, Riyadh, Saudi Arabia
}
\maketitle              % typeset the header of the contribution
\begin{abstract}
Reliable interpretation of cardiac ultrasound images is essential for accurate clinical diagnosis and assessment. Self-supervised learning has shown promise in medical imaging by leveraging large unlabelled datasets to learn meaningful representations. In this study, we evaluate and compare two self-supervised learning frameworks, USF-MAE, developed by our team, and MoCo v3, on the recently introduced CACTUS dataset (37,736 images) for automated simulated cardiac view (A4C, PL, PSAV, PSMV, Random, and SC) classification. Both models used 5-fold cross-validation, enabling robust assessment of generalization performance across multiple random splits. The CACTUS dataset provides expert-annotated cardiac ultrasound images with diverse views. We adopt an identical training protocol for both models to ensure a fair comparison. Both models are configured with a learning rate of 0.0001 and a weight decay of 0.01. For each fold, we record performance metrics including ROC-AUC, accuracy, F1-score, and recall. Our results indicate that USF-MAE consistently outperforms MoCo v3 across metrics. The average testing AUC for USF-MAE is 99.99\% ($\pm$0.01\% 95\% CI), compared to 99.97\% ($\pm$0.01\%) for MoCo v3. USF-MAE achieves a mean testing accuracy of 99.33\% ($\pm$0.18\%), higher than the 98.99\% ($\pm$0.28\%) reported for MoCo v3. Similar trends are observed for the F1-score and recall, with improvements statistically significant across folds (paired t-test, p=0.0048 < 0.01). This proof-of-concept analysis suggests that USF-MAE learns more discriminative features for cardiac view classification than MoCo v3 when applied to this dataset. The enhanced performance across multiple metrics highlights the potential of USF-MAE for improving automated cardiac ultrasound classification. 

\end{abstract}

\section{Introduction}
Cardiac ultrasound (echocardiography) is a cornerstone imaging modality in both adult and fetal cardiology due to its real-time assessment of cardiac anatomy and function, lack of ionizing radiation, and widespread clinical availability. However, reliable interpretation of echocardiographic images requires extensive training and experience, and manual annotation of cardiac views is both time-consuming and subject to inter-observer variability \cite{b1}. The complexity of cardiac ultrasound interpretation is further amplified in fetal imaging, where small cardiac structures and variable fetal positions increase the difficulty of consistent view identification.

Deep learning has demonstrated substantial promise for automating medical image analysis, achieving expert-level performance on diverse tasks across modalities such as radiography, ultrasound \cite{b8,b9,b10,b11,b12,b13,b14}, and magnetic resonance imaging \cite{b2}. Traditional supervised learning approaches rely heavily on large quantities of labelled data, which are often scarce and costly to obtain in clinical settings. In contrast, self-supervised learning (SSL) enables models to leverage vast collections of unlabelled medical images to learn useful representations, subsequently fine-tuning on smaller annotated datasets with improved performance and annotation efficiency \cite{b2,b3}.

Recent trends in SSL include both contrastive methods and generative pretext tasks. Methods such as momentum contrast (MoCo) \cite{b15} exploit a contrastive learning objective to cluster semantically similar examples in embedding space without labels \cite{b4}. Masked autoencoder (MAE) approaches \cite{b5}, originally introduced for natural images, learn to reconstruct masked portions of inputs, effectively encouraging models to capture rich, global structure in image representations. These pre-training strategies have been adapted to medical imaging tasks, and empirical evidence suggests that in-domain SSL pre-training can yield better downstream performance than models initialized on natural images alone \cite{b8,b6}.

\begin{table}
\caption{Dataset distribution of cardiac view classes and data split used in stratified 5-fold cross-validation.}\label{tab:dataset_distribution}
\begin{tabular}{|l|l|}
\hline
\textbf{Category} & \textbf{Number of Images} \\
\hline
Total Images & 37,736 \\
\hline
A4C & 7,422 \\
PL & 6,102 \\
PSAV & 5,832 \\
PSMV & 6,014 \\
Random & 6,021 \\
SC & 6,345 \\
\hline
Testing Set (per fold) & 7,547 \\
Training Set (4-folds) & 30,189 \\
\hline
\end{tabular}
\end{table}

In the domain of cardiac view classification, prior work has applied contrastive SSL to learn discriminative features from echocardiograms, improving classification accuracy relative to purely supervised baselines \cite{b1}. More recently, the CACTUS dataset was introduced as the first open, graded large-scale dataset of cardiac ultrasound views, encompassing multiple standard views and random images to support both classification and quality assessment tasks \cite{b7}. The availability of CACTUS enables rigorous evaluation of self-supervised models tailored specifically for cardiac ultrasound representations.

Despite this progress, there remains a need to systematically benchmark SSL frameworks on CACTUS to identify the most effective pre-training paradigms for cardiac ultrasound view classification. Furthermore, foundation models trained on large, diverse unlabeled ultrasound corpora may offer improved transferability compared with SSL models pre-trained on smaller or natural image datasets. In this study, we evaluate and compare our recently published ultrasound self-supervised foundation model with masked autoencoding (USF-MAE) \cite{b8} with MoCo v3 \cite{b18} on the CACTUS dataset \cite{b7}. This comparison serves as a proof-of-concept (POC) to assess whether large-scale, domain-specific self-supervised pre-training yields more discriminative features than contrastive learning for cardiac view classification, an essential first step towards downstream applications such as congenital heart defects (CHD) detection.

% \subsection*{Contributions}
% The primary contributions of this work are as follows:
% \begin{itemize}
%     \item We perform a controlled comparison between two self-supervised frameworks, USF-MAE and MoCo v3, on the publicly available CACTUS cardiac ultrasound dataset.
%     \item We adopt identical fine-tuning protocols to ensure a fair evaluation and report comprehensive performance metrics, including ROC-AUC, accuracy, F1-score, and recall across 5-fold cross-validation.
%     \item We demonstrate that USF-MAE consistently outperforms MoCo v3 across multiple metrics by a small margin, suggesting the utility of large-scale, ultrasound-specific foundation models for cardiac view understanding.
% \end{itemize}

\section{Methodology}
\subsection{Dataset}
We conducted all experiments using the publicly available CACTUS dataset \cite{b7}, which contains 37,736 expert-annotated cardiac ultrasound images generated by a phantom across six classes: apical four-chamber (A4C), parasternal long-axis (PL), parasternal short-axis aortic valve (PSAV), parasternal short-axis mitral valve (PSMV), Random, and subcostal four-chamber (SC) views (Table~\ref{tab:dataset_distribution}). The Random views class includes non-standard or non-diagnostic frames, increasing classification difficulty and better simulating real-world variability.

\begin{figure}
\includegraphics[width=\textwidth]{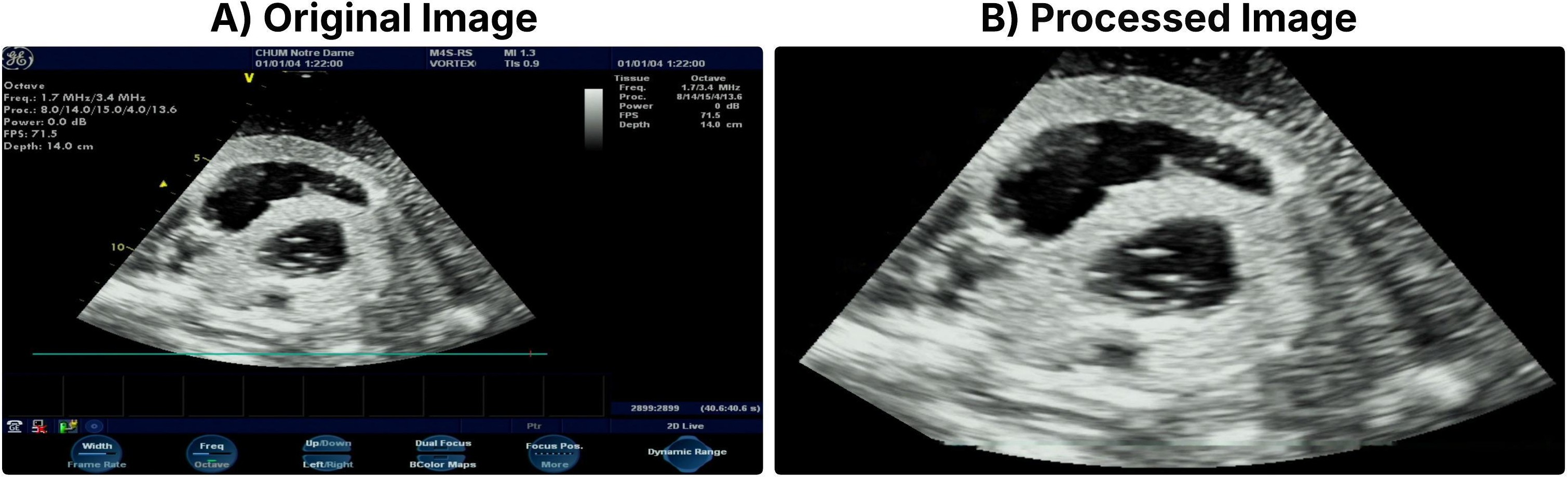}
\caption{A) Original ultrasound frame with overlay annotations. B) Cropped and cleaned, processed and inpainted image used for analysis.} \label{preprocessing_example}
\end{figure}

To ensure robust evaluation, we adopted a stratified 5-fold cross-validation protocol such that each of the six classes was equally represented in every fold. In each fold, four splits were used for training, and the remaining split was used for testing.

\subsection{Image Preprocessing}
Ultrasound images often contain sector-shaped acquisition regions and colour annotations or markers, as shown in Fig.~\ref{preprocessing_example}A. To standardize the input and reduce irrelevant visual artifacts, we applied a three-stage preprocessing pipeline.

\textbf{Sector masking and cropping.}  
Each image was first converted to RGB, and a sector mask was applied to isolate the ultrasound field of view. The mask was defined using a pie-slice geometry centred at the bottom midpoint of the image with angular limits of $210^\circ$ to $330^\circ$ and radius equal to 90\% of the image height. Pixels outside this region were set to zero. The masked image was then cropped using fixed bounding coordinates to remove peripheral borders and scanner overlays.

\textbf{Annotation mask extraction.}  
To remove embedded colour annotations, images were converted to the HSV colour space. Colour thresholds were applied to detect yellow, blue, and red overlays commonly used for measurement markers. Binary masks were generated using these thresholds and subsequently dilated with a $5 \times 5$ kernel to ensure complete coverage of annotation regions.

\textbf{Inpainting.}  
Detected annotation regions were removed and filled in the missing pixel values using the Navier–Stokes–based inpainting \cite{b16} implemented in OpenCV (Fig.~\ref{preprocessing_example}B). This step ensured that classification performance reflected anatomical content rather than overlaid measurement graphics or noisy artifacts.

\begin{table}[t]
\centering
\caption{Comparison of pre-training configurations for MoCo v3 and USF-MAE.}
\begin{tabular}{|p{3.6cm}|p{4.1cm}|p{4.2cm}|}
\hline
\textbf{Component} & \textbf{MoCo v3 \cite{b18}} & \textbf{USF-MAE (ours) \cite{b8}} \\
\hline
Backbone architecture & ViT-B/16 & ViT-B/16 \\
\hline
Pretraining paradigm & Contrastive SSL & Masked autoencoding (MAE) \\
\hline
Pretraining dataset & ImageNet-1K & OpenUS-46 \\
\hline
Dataset domain & Natural images & Ultrasound images \\
\hline
Number of images & $\sim$1.28M & $\sim$370K \\
\hline
Pretraining objective & Instance discrimination via momentum contrast & Patch reconstruction (MSE over masked patches) \\
\hline
Masking ratio & N/A & 25\% \\
\hline
Pretrained weights & Publicly released weights & Publicly released weights \\
\hline
Fine-tuning on CACTUS & Full model fine-tuning & Full model fine-tuning \\
\hline
\end{tabular}
\label{tab:pretraining_comparison}
\end{table}

\subsection{Model Architectures}
We benchmarked two self-supervised frameworks: MoCo v3 \cite{b18}, and our previously developed ultrasound foundation model, USF-MAE \cite{b8}. A detailed comparison of their pretraining configurations is provided in Table~\ref{tab:pretraining_comparison}.

\subsubsection{MoCo v3 Fine-Tuning}
For MoCo v3 \cite{b18}, we initialized a Vision Transformer (ViT)-base model (ViT-B/16) backbone with publicly available MoCo v3 SSL pretrained weights. The classification head was replaced with a linear layer mapping the ViT feature dimension to six output classes. All backbone parameters were fine-tuned during training.

\begin{figure*}[t]
\centering
\includegraphics[width=1\textwidth]{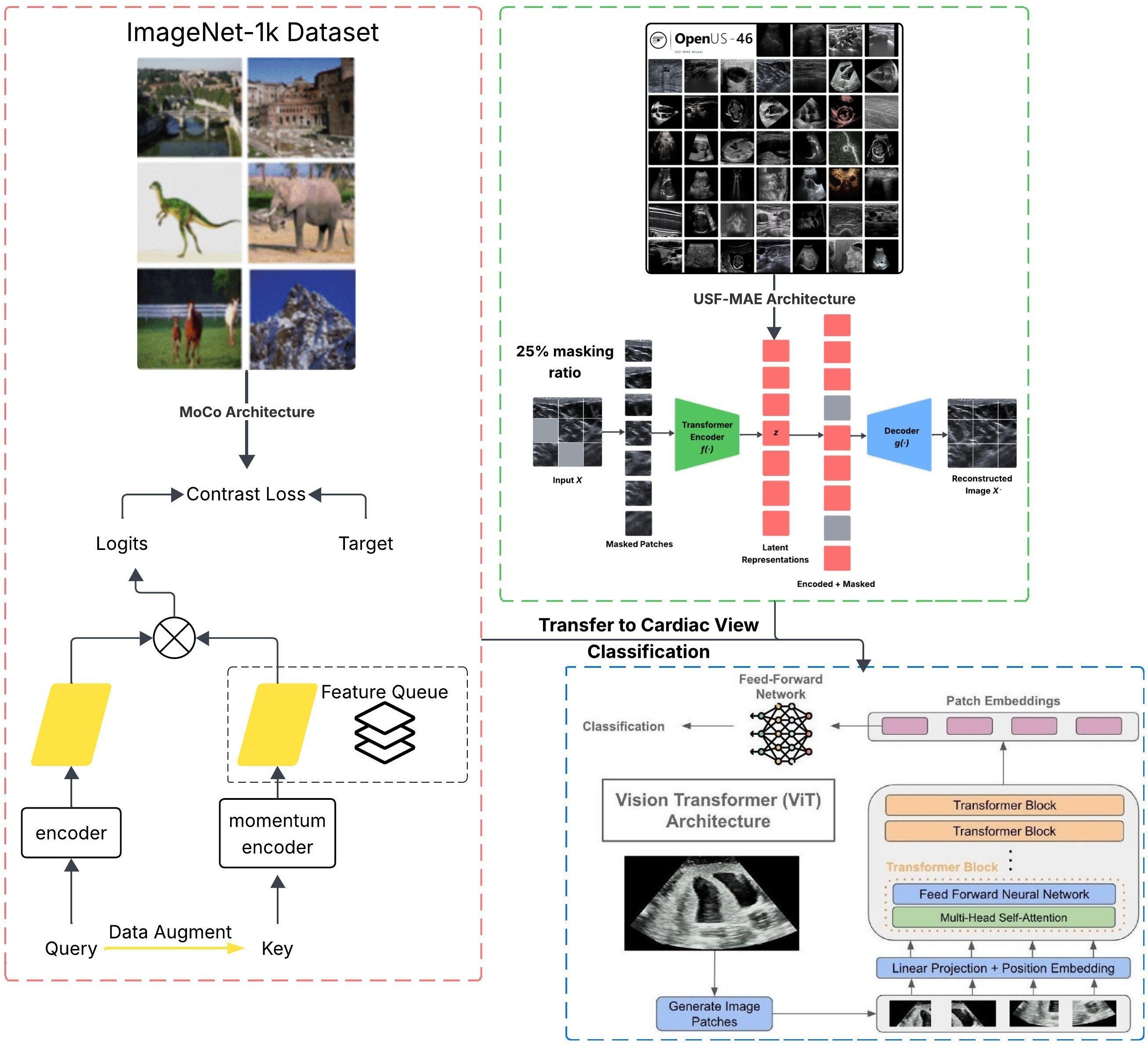}
\caption{Comparison of MoCo v3 and USF-MAE Pipelines.} \label{model_arc}
\end{figure*}

\subsubsection{USF-MAE Fine-Tuning}
For USF-MAE \cite{b8}, we initialized its ViT-B/16 backbone with our publicly available ultrasound-specific USF-MAE SSL pretrained weights in GitHub, similar to MoCo v3. The classification head was replaced with a linear layer mapping the ViT feature dimension to six output classes. All backbone parameters were fine-tuned during training as well. Consequently, the two models share identical backbone architecture and fine-tuning protocol, differing only in their pretraining strategy and data domain, namely natural images versus ultrasound images (Table~\ref{tab:pretraining_comparison}).

% \subsubsection{USF-MAE Fine-Tuning}
% USF-MAE \cite{b8} is an MAE pretrained on large-scale ultrasound data (OpenUS-46: $\sim$370,000 images) using a reconstruction objective by masking 25\% of input image patches and learning how to reconstruct them (Fig.~\ref{model_arc}). Let $x$ denote an input image divided into non-overlapping patches, and let $\mathcal{M}$ denote the set of masked patches. The encoder processes only the visible patches, and a lightweight decoder predicts the pixel values of the masked patches. The pretraining objective minimizes the mean squared reconstruction error over masked patches:

% \begin{equation}
% \mathcal{L}_{\text{MAE}} 
% = \frac{1}{|\mathcal{M}|} 
% \sum_{i \in \mathcal{M}} 
% \left\| x_i - \hat{x}_i \right\|_2^2,
% \end{equation}

% where $x_i$ and $\hat{x}_i$ represent the original and reconstructed pixel values of the $i$-th masked patch, respectively. 

% The encoder architecture is based on ViT-B/16. For this study, the pretrained encoder was transferred to the CACTUS dataset, and a linear classification head was appended. The full network was fine-tuned end-to-end under the same training configuration used for MoCo v3 to ensure a fair comparison.

\subsection{Training and Testing Protocol}
To ensure comparability, both models were trained using identical hyperparameters and optimization strategies (end-to-end fine-tuning). All experiments were conducted using NVIDIA L40S GPU acceleration.

\textbf{Input processing.}  
Images were resized to $224 \times 224$ pixels. During training, data augmentation included random rotation (0–90$^\circ$), horizontal and vertical flipping of 50\% probability, and random resized cropping with scale range [0.5, 2.0]. Pixel intensities were normalized using the mean of [0.485, 0.456, 0.406] and the standard deviation of [0.229, 0.224, 0.225].

\textbf{Optimization.}  
Training was performed for 15 epochs per fold using AdamW \cite{b17} with a learning rate of 0.0001 and a weight decay of 0.01. A cosine learning rate scheduler with linear warm-up was applied. The batch size was set to 32. A weighted cross-entropy loss was employed. Class weights were computed from the training split of each fold using inverse frequency balancing.

\textbf{Model selection.}  
For each fold, the model achieving the lowest training loss was selected as the best checkpoint.

\textbf{Evaluation Metrics.} 
Performance was evaluated on the testing split of each fold after training and averaged across all five folds. Reported metrics include accuracy, weighted recall, weighted F1-score, and macro one-versus-rest ROC-AUC. This experimental framework allows a controlled assessment of representation quality learned through different self-supervised pretraining paradigms when transferred to cardiac ultrasound view classification.

% \begin{table}
% \caption{Performance comparison (\%) of cardiac view classification models on the mean testing sets across 5-fold cross-validation.}\label{results_table}
% \begin{tabular}{|l|l|l|l|l|}
% \hline
% \textbf{Model} & \textbf{Accuracy} & \textbf{Recall} & \textbf{F1-score} & \textbf{AUC} \\
% \hline
% MoCo v3 & 98.99$\pm$0.0011 & 98.99$\pm$0.0011 & 98.99$\pm$0.0011 & 99.97$\pm$0.00006\\
% \textbf{USF-MAE (ours)} & \textbf{99.33$\pm$0.0007} & \textbf{99.33$\pm$0.0007} & \textbf{99.33$\pm$0.0007} & \textbf{99.99$\pm$0.00004}\\
% \hline
% \end{tabular}
% \end{table}

\section{Results}
Table~\ref{results_table} summarizes the mean testing performance across 5-fold cross-validation for both self-supervised frameworks. Overall, both models achieved near-perfect discrimination on the CACTUS dataset; however, USF-MAE consistently outperformed MoCo v3 across all evaluation metrics.

USF-MAE achieved a mean testing accuracy of 99.33\% ($\pm$0.18\% 95\% CI), compared to 98.99\% ($\pm$0.28\% 95\% CI) for MoCo v3. A similar improvement was observed in weighted F1-score (99.33\% vs. 98.99\%) and recall (99.33\% vs. 98.99\%). In terms of macro ROC-AUC, USF-MAE reached 99.99\% ($\pm$0.01\% 95\% CI), slightly higher than the 99.97\% ($\pm$0.01\% 95\% CI) obtained with MoCo v3. To assess whether the observed performance differences were statistically significant, we conducted a paired t-test on the fold-wise F1-scores. USF-MAE achieved significantly higher F1-scores than MoCo v3 across the five folds, p=0.0048 (< $\alpha$ of 0.01), corresponding to a mean improvement of 0.34\% points.

\begin{figure}
\includegraphics[width=\textwidth]{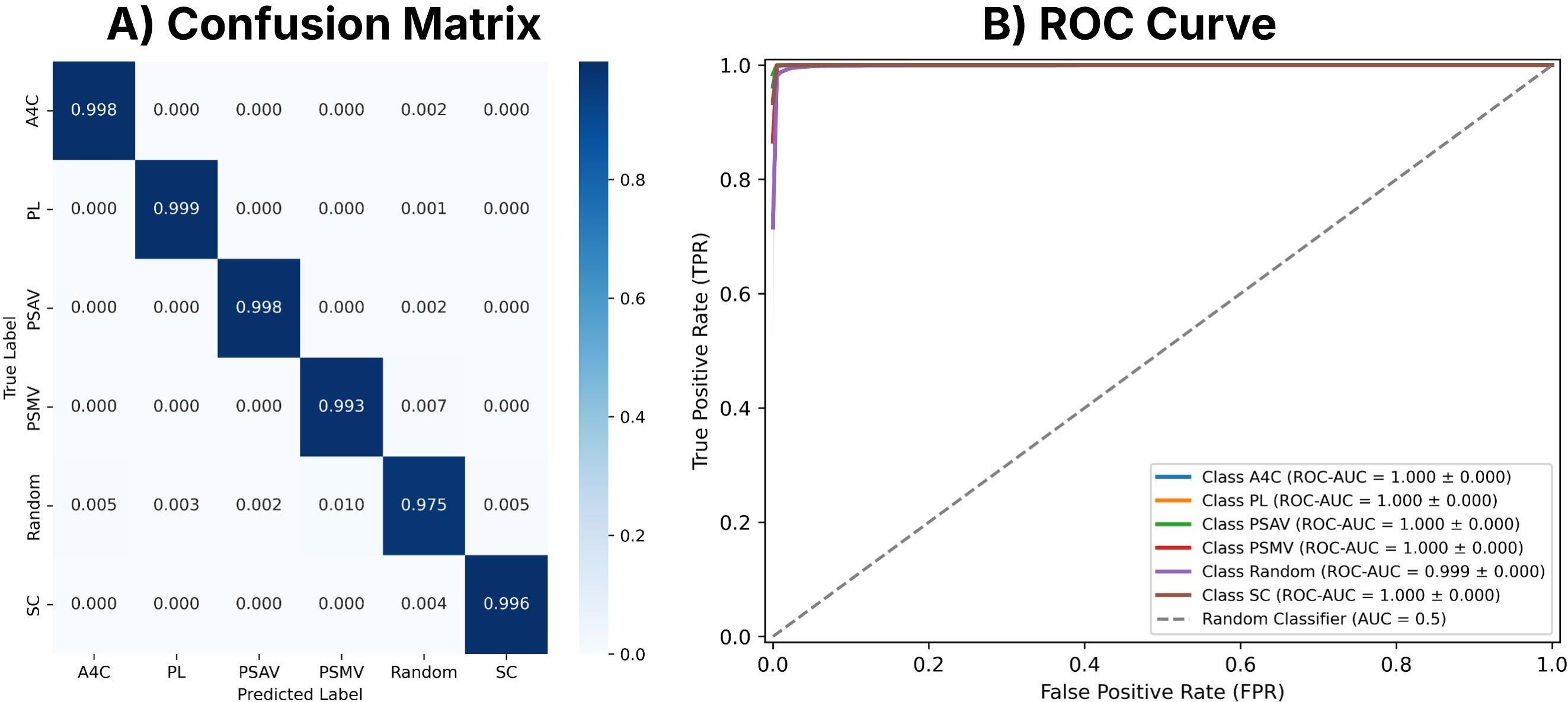}
\caption{Classification performance for cardiac view classification: A) normalized confusion matrix and B) per-class ROC curves showing near perfect discrimination across all classes.} \label{results_figure}
\end{figure}

The normalized confusion matrix for the best-performing USF-MAE model (Fig.~\ref{results_figure}A) demonstrates minimal inter-class confusion, with per-class sensitivities exceeding 97.5\% across all views. The Random class showed slightly higher variability relative to standard anatomical views, but remained highly discriminative. 

The per-class ROC curves (Fig.~\ref{results_figure}B) further confirm the strong separability of all five cardiac views and the random class, with AUC values approaching 1.0 for each class. These findings indicate that both self-supervised approaches learn highly discriminative representations for cardiac view classification, with USF-MAE providing consistently superior performance under identical fine-tuning conditions.

Collectively, these results support the hypothesis that large-scale ultrasound-specific MAE pretraining yields more transferable and discriminative features than contrastive pretraining when applied to cardiac ultrasound view classification, within the scope of this POC study.

\begin{table}[t]
\centering
\small   % or \footnotesize or \scriptsize
\setlength{\tabcolsep}{3pt}
\caption{Per-fold performance across 5-fold cross-validation.}
\label{results_table}
\begin{tabular}{|c|cc|cc|cc|cc|}
\hline
\multirow{2}{*}{\textbf{Fold}} 
& \multicolumn{2}{c|}{\textbf{Accuracy (\%)}} 
& \multicolumn{2}{c|}{\textbf{Recall (\%)}} 
& \multicolumn{2}{c|}{\textbf{F1-score (\%)}} 
& \multicolumn{2}{c|}{\textbf{AUC (\%)}} \\
\cline{2-9}
& MoCo v3 & \textbf{USF-MAE} & MoCo v3 & \textbf{USF-MAE} & MoCo v3 & \textbf{USF-MAE} & MoCo v3 & \textbf{USF-MAE} \\
\hline

1 & 99.18 & \textbf{99.35} & 99.18 & \textbf{99.35} & 99.18 & \textbf{99.35} & 99.98 & \textbf{99.99} \\
2 & 99.01 & \textbf{99.35} & 99.01 & \textbf{99.35} & 99.01 & \textbf{99.35} & 99.97 & \textbf{99.99} \\
3 & 98.91 & \textbf{99.42} & 98.91 & \textbf{99.42} & 98.91 & \textbf{99.42} & 99.97 & \textbf{99.98} \\
4 & 98.97 & \textbf{99.22} & 98.97 & \textbf{99.22} & 98.97 & \textbf{99.22} & 99.98 & \textbf{99.99} \\
5 & 98.90 & \textbf{99.31} & 98.90 & \textbf{99.31} & 98.90 & \textbf{99.31} & 99.98 & \textbf{99.98} \\
\hline
Mean & 98.99 & \textbf{99.33} & 98.99 & \textbf{99.33} & 98.99 & \textbf{99.33} & 99.97 & \textbf{99.99} \\
95\% CI & 0.28 & \textbf{0.18} & 0.28 & \textbf{0.18} & 0.28 & \textbf{0.18} & \textbf{0.01} & \textbf{0.01} \\
\hline
\end{tabular}
\end{table}

\section{Discussion}
In this POC study, we benchmarked two SSL paradigms, contrastive learning (MoCo v3) and masked autoencoding (USF-MAE), for cardiac ultrasound view classification on the CACTUS dataset. While both approaches achieved near-ceiling performance across all evaluation metrics, USF-MAE consistently demonstrated superior accuracy, F1-score, and ROC-AUC under identical fine-tuning conditions. These results suggest that ultrasound-specific MAE pretraining produces highly transferable representations for cardiac view discrimination.

Two key factors differentiate the evaluated models: (1) the SSL objective (masked autoencoding vs. contrastive learning), and (2) the pretraining data domain (ultrasound-specific data vs. natural images). MoCo v3 was pretrained using contrastive learning on ImageNet-scale natural images, whereas USF-MAE was pretrained using masked autoencoding on large-scale ultrasound data. While both methodological differences may contribute to performance variation, prior studies have consistently demonstrated that domain-specific pretraining often has a larger impact on downstream medical imaging performance than initialization from natural image datasets alone. For example, Raghu et al. \cite{b19} showed that ImageNet representations may not fully transfer to medical imaging tasks due to domain mismatch. Similarly, Azizi et al. \cite{b20} demonstrated that large-scale self-supervised pretraining within the medical domain substantially improves downstream classification performance compared to natural image supervised pretraining.

% Although the absolute gain in accuracy, as an example, is 0.34\%, this corresponds to a substantial reduction in classification error. Specifically, the error rate decreased from 1.01\% to 0.67\%, representing a relative error reduction of 33.7\%. In already high-performance regimes, such reductions indicate meaningful improvements in representation quality rather than marginal numerical gains. Compared to MoCo v3, which is pretrained using instance-level contrastive learning on natural images, USF-MAE is pretrained using masked autoencoding on large-scale ultrasound data. Masked autoencoding encourages reconstruction of structurally coherent regions, requiring the encoder to model global anatomical context and spatial dependencies. In contrast, contrastive learning primarily optimizes instance discrimination, which may emphasize feature separability rather than detailed anatomical structure. Given the importance of spatial relationships and organ-level context in ultrasound, this pretraining strategy may provide USF-MAE with a domain-specific advantage over MoCo v3.

Although the absolute gain in accuracy, as an example, is 0.34\%, this corresponds to a substantial reduction in classification error. Specifically, the error rate decreased from 1.01\% to 0.67\%, representing a relative error reduction of 33.7\%. In already high-performance regimes, such reductions indicate meaningful improvements in representation quality rather than marginal numerical gains. Given existing evidence supporting the importance of domain alignment in medical imaging, we hypothesize that ultrasound-specific pretraining contributes more substantially to the observed performance gain than the difference in SSL objective alone.

Importantly, MoCo v3 also achieved excellent performance, confirming that contrastive SSL remains a strong baseline for cardiac ultrasound representation learning. The near-perfect discrimination observed for both models suggests that CACTUS view classification is a suitable benchmarking task for evaluating representation quality before using the models for more complex downstream tasks.

From a clinical perspective, accurate cardiac view classification is a foundational prerequisite for automated CHD detection in fetal echocardiography. CHD diagnosis depends heavily on the correct acquisition and identification of standard views before structural abnormalities can be assessed. Therefore, robust and transferable representations learned through self-supervised pretraining may facilitate improved generalization when transitioning from view classification to CHD detection tasks.

Several limitations should be acknowledged. First, the CACTUS dataset consists of simulator-acquired images generated by scanning a phantom, which may not fully capture the variability present in real-world ultrasound examinations. Second, performance was evaluated on a single dataset, and external validation was not performed in this study. Although this study focused on a single dataset, evidence from our recent work on first-trimester fetal heart view classification \cite{b12} supports the generalizability of ultrasound-specific SSL. In that study, USF-MAE also outperformed all baselines on real fetal ultrasound images, demonstrating the practical utility of this pretraining paradigm beyond simulator data. Third, the classification task is inherently easier than fine-grained abnormality detection, and performance differences may become more pronounced in more challenging downstream tasks.

Future work will extend this analysis beyond view classification to grading and diagnostic performance using the CACTUS dataset, enabling evaluation of whether ultrasound-specific pretraining improves clinically relevant assessment tasks. We also plan to validate the models on real fetal echocardiography datasets and examine the impact of ultrasound-specific foundation pretraining on CHD detection and multi-class diagnostic classification in the near future. Such investigations will determine whether the observed representation advantages translate into meaningful clinical gains.

% Overall, this study provides empirical evidence supporting the use of large-scale ultrasound-specific MAE pretraining as a strong foundation for cardiac ultrasound representation learning.

\section{Conclusion}
% In this study, we benchmarked two SSL frameworks, USF-MAE and MoCo v3, for cardiac ultrasound view classification on the CACTUS dataset. Under identical fine-tuning conditions, USF-MAE consistently achieved superior performance across accuracy, recall, F1-score, and ROC-AUC metrics. These findings suggest that large-scale ultrasound-specific MAE pretraining yields highly transferable representations for cardiac view discrimination. The USF-MAE framework, along with its pretrained weights, can be accessed publicly through our GitHub repository: https://github.com/Yusufii9/USF-MAE. As a POC analysis, this work supports the use of ultrasound foundation models as a strong initialization strategy for downstream clinical tasks. Future studies will investigate whether these representation advantages translate to improved detection of CHD in real fetal echocardiography settings.

In this study, we benchmarked USF-MAE and MoCo v3 for cardiac ultrasound view classification on the CACTUS dataset. Under identical fine-tuning conditions, USF-MAE achieved consistently superior performance across all evaluation metrics. These findings suggest that ultrasound-specific MAE pretraining provides highly transferable representations for cardiac view discrimination. The USF-MAE framework and pretrained weights are publicly available at: https://github.com/Yusufii9/USF-MAE. As a POC analysis, this work supports ultrasound foundation models as strong initializations for downstream clinical tasks. Future studies will evaluate whether these advantages translate to improved CHD detection in real fetal echocardiography.

\section*{Use of AI Assistance}
The author(s) used ChatGPT (by OpenAI) for language editing (grammar) and sentence clarity improvement. The tool was used solely to enhance readability and refine sentence structure. It was not used to generate scientific content, experimental results, data analysis, or technical conclusions. After using this tool, the author(s) carefully reviewed and edited the output and take full responsibility for the contents of the manuscript.

\end{document}